**Assessing The Spatially Heterogeneous Impact of Recurrent Flooding On Accessibility: A Case Study of The Hampton Roads Region – Part 2: Transit Accessibility**

**Luwei Zeng**
Graduate Research Assistant
Department of Engineering System and Environment
University of Virginia, Charlottesville, VA, 22903
lz6ct@virginia.edu

**T. Donna Chen, PE, Ph.D.**
Corresponding Author
Assistant Professor
Department of Engineering Systems and Environment
University of Virginia, Charlottesville, VA 22903
tdchen@virginia.edu

**John S. Miller, PE, Ph.D.**
Research Scientist
Virginia Transportation Research Council
Charlottesville, VA, 22903
John.Miller@VDOT.Virginia.gov

**Jonathan L. Goodall**
Professor
Department of Engineering, Systems and Environment
University of Virginia, Charlottesville, VA 22903
goodall@virginia.edu

**Faria Tuz Zahura**
Graduate Research Assistant
Department of Engineering Systems and Environment
University of Virginia, Charlottesville, VA 22903
Fz7xb@virginia.edu

**ABSTRACT**

Due to accelerated sea level rise and climate change, the transportation system is increasingly affected by recurrent flooding coastal regions, yet the cumulative travel disruption effects are not well understood. In Part 1 of this study, the accessibility impacts of recurrent flooding on the auto mode were examined. In this paper (Part 2 of the study), the impact of recurrent flooding on transit service accessibility was quantified with the aid of spatially and temporally disaggregated crowdsourced flood incident data from WAZE. A fixed route transit network is built for five time-of-day periods for 710 traffic analysis zones (TAZs), to capture the spatial and temporal variation of transit accessibility reduction due to recurrent flooding. Results show that the greatest transit accessibility reduction occurs during the morning peak hour, with individual TAZ transit accessibility reduction ranging from 0 to 88.2% for work trips (with an average of 6.4%) and ranging from 0 to 99.9% for non-work trips (with an average of 3.7%). Furthermore, social vulnerability analysis indicates that TAZs with a greater share of people with higher vulnerability in transportation and socio-economic status are more likely to experience recurrent flooding-induced transit accessibility reduction. Results from this study reinforce the notion that transportation impacts under recurrent flooding are not uniformly experienced throughout a region, and this spatial and temporal variation translates to different impacts borne by various population groups. Disaggregate impact analysis like this study can support transportation engineers and planners to prioritize resources to ensure equitable transit accessibility under increasing climate disruptions.

## 1. INTRODUCTION

Recurrent flooding, or nuisance flooding, has become an increasingly common disruption to the transportation system due to climate change and sea level rise (*1*). Recurrent flooding is considered a small-scale climate event compare to disaster flooding. Recurrent flooding has localized characteristic where incidents often occur at repeated locations, but attracts less attention compared to large-scale flooding events (*2*, *3*). However, the localized nature of recurrent flooding implies that specific subgroups of travelers are repeatedly affected by these climate events, and the increasing cumulative impact should not be ignored.

Part 1 of the study investigated recurrent flooding's impacts on automobile accessibility and found that socioeconomically vulnerable travelers are disproportionately affected. However, socioeconomically vulnerable travelers are less likely to have access to private automobile travel. Hence, this paper (Part 2 of the study) focuses on demonstrating a framework to assess the impacts of recurrent flooding on transit service accessibility, including which populations are more likely to be impacted. Previous studies have shown the socially vulnerable to be more at risk for climate-induced hazards (*4*, *5*). Furthermore, various transportation and infrastructure systems might have various degrees of tolerance to such climate events (*4*, *5*). This study focuses on transit accessibility variation under recurrent flooding, because fixed route transit services cannot detour as freely as private vehicles and might experience higher impact under recurrent roadway flooding. The relative inflexibility of fixed route transit also implies that recurrent flooding's spatially heterogeneous impacts might be further exacerbated in specific locations within the transit service network. Understanding the spatial distribution and extent of recurrent flooding impact is critical for transportation infrastructure investment decisions, as well as improving social equity for those most affected by these impacts.

## 2. PURPOSE

The aim of this paper is to examine the impact of recurrent flooding on fixed route transit accessibility in the coastal Hampton Roads region in Virginia, USA. The research objectives are to (1) developed a framework to quantify the impact of recurrent flooding on transit network with the aid of crowdsourced data, (2) demonstrate the methodology in a case study region with publicly accessible data, and (3) identify the specific areas that are most impacted by recurrent flooding and any over-representation of socially vulnerable populations in those areas. Figure 1(a) shows the entire Hampton Roads region and Figure 1(b) shows the area with public transit coverage within the region, the analysis area of this study. These selected traffic analysis zones (TAZs) in Figure 1(b) fall within a 0.25 mile radius (*6*) of a transit stop, thereby they are considered accessible via transit service.

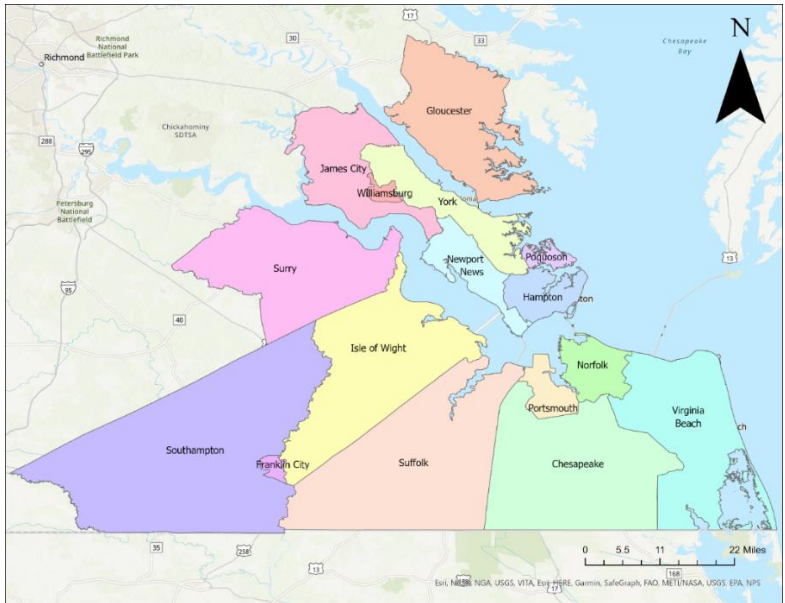

**Figure 1 (a) Hampton Roads Region, Virginia**

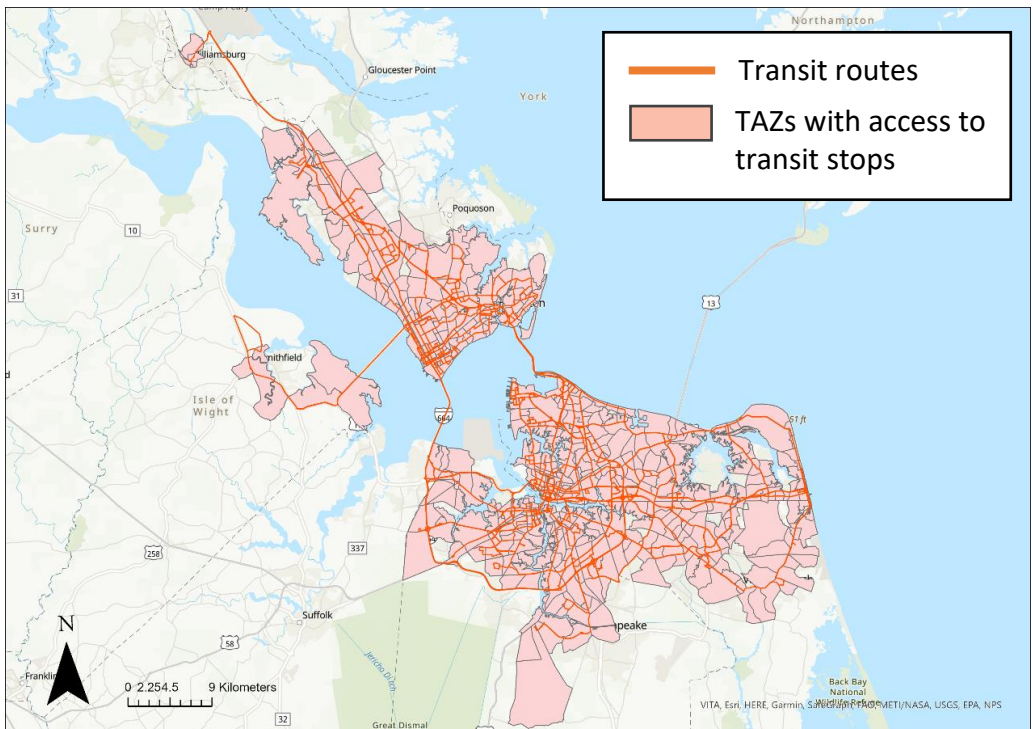

**Figure 1(b) Case Study Area: Hampton Roads traffic analysis zones with transit access**

## 3. LITERATURE REVIEW

As studies on transportation system impacts of small- and large-scale climate events have been reviewed and discussed in Part 1 of this study, only a short summary of those studies are presented here. The impacts of flooding has been studied by existing literature (*7–19*) as the frequency of flooding events have increased, including general trends in flooding events, the impacted population and lands, and various methodological approaches of quantifying flooding impacts. With sea level rise, flooding has been shown to affect 20% of the population's access to critical infrastructures in Boston (*11*), and lead to inundation in homes and damaged dwelling units (*20*). Simulation methods with the aid of climate models are widely used for understanding the impact of climate events on transportation infrastructure (*7–16*). Combined with traffic models, these studies are able to quantify the impact of climate events with mobility indicators such as change in travel time and free flow speeds (*21–23*). Studies of the impact of major flooding events (caused by extreme rainfall or storm surges) on the transport network are enhanced by incorporating GIS network analysis and other numerical modeling methods (*7–19*). The consequences of disastrous climate events are long-lasting and widely noticed, thus existing studies are largely focused on such events.

As the focus of this paper is on transit accessibility, the remainder of the literature review is devoted to 1) methods from existing studies quantifying transit accessibility, 2) results from existing studies focused on transit accessibility impacts of climate events, and 3) previous studies that study the relationship between social vulnerability and transit accessibility.

Previous literature on transit accessibility measurements are categorized into three models by Malekzadeh et al. (*24*): system accessibility (measure the ease of people's access to transit stops), system-facilitated accessibility (measure the cost of people's access to a single destination via transit using the metric of travel time), and integral accessibility (measure the ease of people's access to multiple possible destinations). For example, under the first category, Foda & Osman (*25*) used Euclidean distances along the transit network to measure the walk accessibility to transit stops. Rao et al.(*26*) defined an environmental transit accessibility index model with utility measurement to understand the transit accessibility by walking and cycling to transit stops. As for system-facilitated accessibility, Tribby et al.(*27*) created a high-resolution spatial-temporal public transit model and incorporated socio-economic data to understand the accessibility improvement of walk time, wait time, and in vehicle travel time with new bus routes. Nassir et al.(*28, 29*) showed the importance of considering public transport network characteristics (e.g., in transit time) by incorporating the traveler decision-making process and stop choice behaviors in transit accessibility measures. The third model, integral accessibility, which is also the model used in this study, is widely used in transit accessibility measurement. For example, El-Geneidy et al.(*30*) developed a model including travel time and fares to measure the transit accessibility to job opportunities. Bocarejo et al.(*31*) created a gravity model to measure transit accessibility to jobs between origin-destination pairs, where the impedance function was developed with travel time and affordability index. Similarly, there are studies measuring transit accessibility to jobs and services, with the indicators of walking time, waiting time, and transit fares (*32–34*). Within the aforementioned studies, transit accessibility measures are static and not sensitive to network disruption.

There are studies investigating the effect of flooding on the transportation system as previously discussed. However, most are not focused on public transit systems, rather choosing to evaluate the impacts across the entire roadway network. A small body of studies have explored the impact of climate events on public transit service. Stoica-Fuchs (*35*) assessed the vulnerability of roads and railroad network under 10-year and 100-year flooding events with GIS analysis, and identified the susceptible sectors. Abad et al.(*36*) collected employee data through questionnaires to understand commuting behavior adaptation under flooding events. A study in Wuhan, China by Hong et al. (*37*) showed that transit accessibility of residential communities for urban commuting and education trip purposes are more vulnerable than other trips under rainstorm-induced flooding events. Abad et al.(*38*) analyzed travelers' risk perception of public transit service under flooding events in The Philippines, with model results suggesting that most travelers do not change their decision to take public transit during flooding.

In many parts of the US, transit service is utilized more by socially vulnerable populations who do not have easy access to private vehicles (*30*). Therefore, it is also important to understand the socioeconomic characteristics of the population whose transit accessibility is impacted by recurrent flooding. The Centers for Disease Control and Prevention (CDC) provides a guide for quantifying a social vulnerability index (SVI) and identifying communities that will most likely be vulnerable to hazardous events and require support (*39*). Few studies have incorporated social vulnerability in studying the transit system, but none specifically in the context of climate events. For example, Boisjoly et al.(*40*) showed that accessibility to employment by transit was not equally distributed across the study region, since employment centers were largely clustered in the core region and high-income areas. Similarly, Deboosere et al. (*41*) developed a methodology to measure the transit accessibility to low-income jobs for vulnerable populations. In terms of access to critical health service, Boisjoly et al. (*42*) quantified the static accessibility to hospitals via transit network and related it with socio-economic characteristics. The author emphasized a major limitation of the study is assuming a fixed time for departures, however in reality "individuals may need to visit the hospital at any time of the day." This finding also reinforces the importance of including temporal variation in analyzing accessibility impacts by short-term disruptions such as recurrent flooding.

In summary, the potential for recurrent flooding to hinder transit accessibility is non-negligible. Cumulative access to potential destinations via public transit is an important metric in transportation system evaluation, and the impact of recurrent flooding on transit accessibility has not yet been quantified. This paper fills this research gap by capturing the spatial and temporal variation of transit accessibility change under recurrent flooding. The study also identifies the socially vulnerable populations who more likely to experience these accessibility impacts.

## 4. DATA SOURCES AND PRE-PROCESSING
This section introduces the data sources for the analysis and discusses the pre-processing procedures. Maps throughout this paper were created using by the researchers using ArcGIS® software by Esri[1]. Except for the transit network data, other sets of data remained the same as Part 1, which contains more detailed discussions of these data sets.

### 4.1 Transit Network Data
Transit networks for five time-of-day periods were constructed using a combination of two data sets. The Hampton Roads Transportation Planning Organization (HRTPO) maintains a GIS shapefile of population and employment for each of the 1,173 TAZs in the region (*43*). Using the most recent (2015) shapefiles, the TAZ is the spatial unit of measure for accessibility in this analysis. The transit service data comes from GTFS (General Transit Feed Specification) data of the Hampton Roads Transit (HTR) service, which is collected by the transitland website (*44*). For the study period of August 2018, GTFS data provides the transit schedule, stops, and trips (route) information.

### 4.2 Crowdsourced Points-of-Interest (POI) Data

---

[1] ArcGIS® and ArcMap™ are the intellectual property of Esri and are used herein under license. Copyright © Esri. All rights reserved. For more information about Esri® software, please visit www.esri.com.

A total of transit-accessible 1,421 POIs—not including other employment sites—were extracted from the Open Street Maps (OSM) applications programming interface (API) Overpass using the Python scripting language. Figure 2 shows the locations of points of interest in Hampton Roads transit service area.

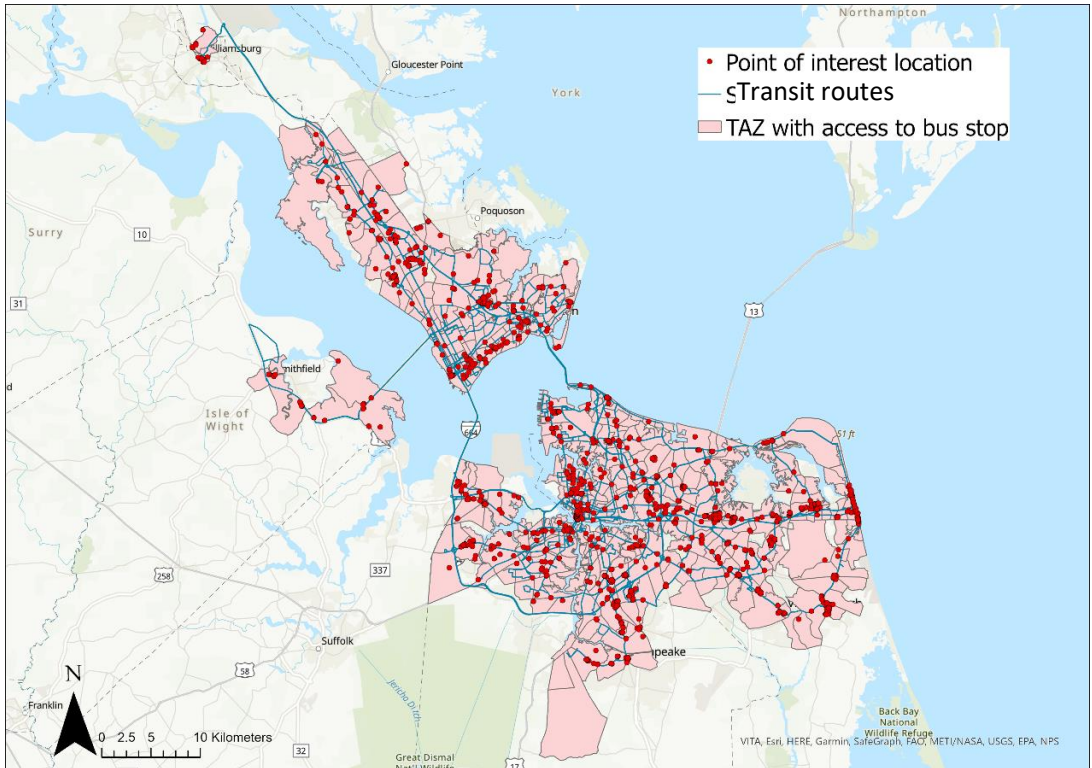

**Figure 2 Locations of Points of Interest in Hampton Roads transit servicing area**

### 4.3 Crowdsourced Flood Incident and Traffic Data

The original data from WAZE is the same as Part 1 of the study, and the pre-processing for selecting valid flooding report remains the same with the adaptation of the transit accessible spatial region. Flood reports were only selected if they were within the transit service areas in this paper.

Table 1 shows the number of flood reports before and after verification within the transit accessible TAZs for each time period across eight days in August 2018. Figure 3 shows the locations of all flood reports for all time-of-day periods in red and the locations of jam data-verified flood reports in green. Based on the verified flood incident reports, there are total of 18 unique time period-day combinations in the analysis.

**Table 1 Verified Flood Incident Reports per Time Period**

| Time period | # Original reports | # Verified reports | % Verified reports |
|---|---|---|---|
| Period 1 (12:00am to 6:00am) | 8 | 4 | 50.0% |
| Period 2 (6:00am to 9:00am) | 51 | 19 | 37.3% |
| Period 3 (9:00am to 3:00pm) | 48 | 20 | 41.7% |
| Period 4 (3:00pm to 6:00pm) | 143 | 110 | 76.9% |

| Period 5 (6:00pm to 12:00am) | 86 | 44 | 51.2% |
|---|---|---|---|

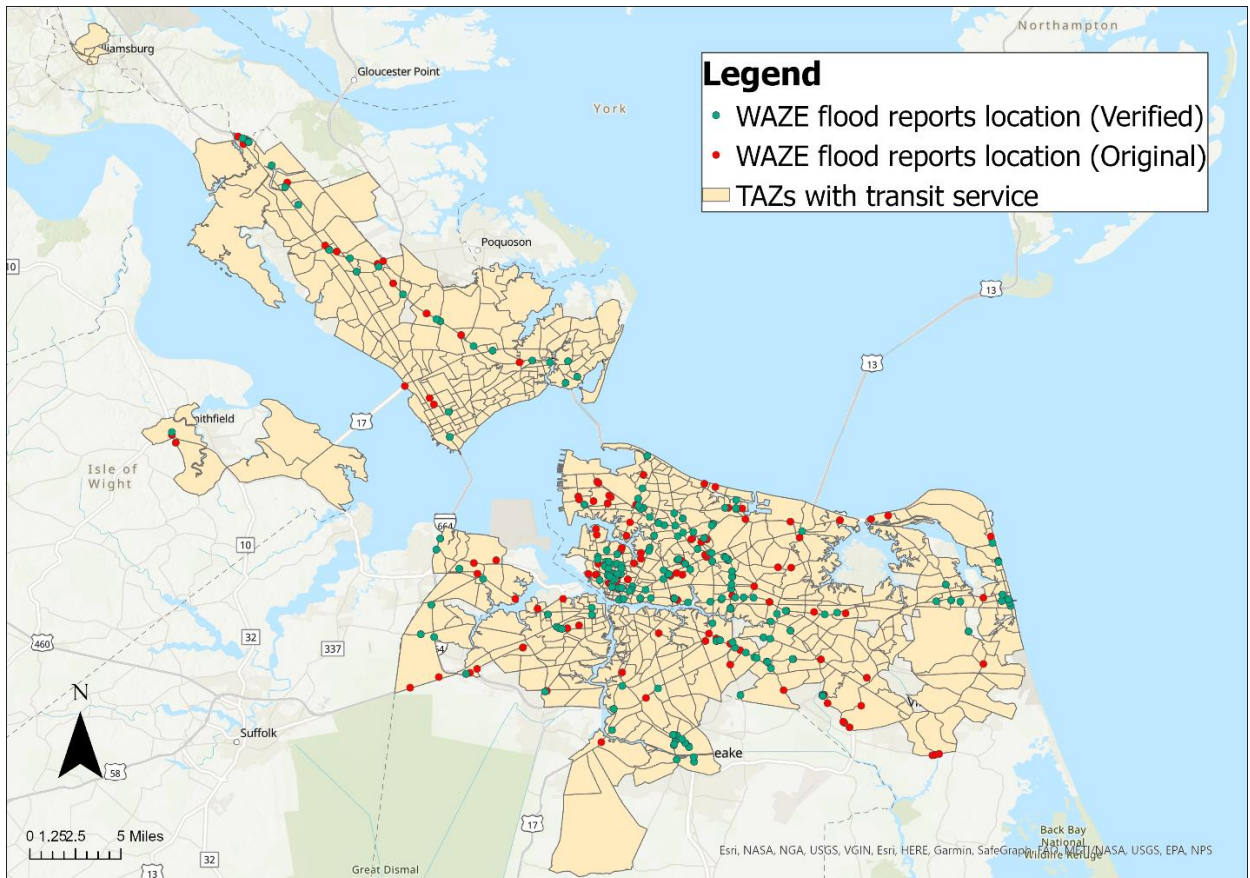

**Figure 3 Locations of WAZE flood reports in transit accessible zones**

## 4.4 Socioeconomic Data

The socioeconomic data remain the same as Part 1 of the study, adapted for the transit accessible zones and collected from the 2017 American Community Survey at the census tract level.

## 5. METHODOLOGY

### 5.1 Transit Network Construction

Based on the GTFS data, each transit route is broken into segments by the stops. The travel time of each segment is calculated based on the GTFS-provided schedule. The service schedule varies throughout the day (e.g. peak commute times offers the largest transit network and overnight period has the most restricted transit network), thereby five transit networks were built to reflect varying levels of transit service across five time-of-day periods. Moreover, the transit network is separated into inbound and outbound networks in order to be compatible with the network analysis tool in ArcGIS Pro, thus each bus stop is also separated into two (co-located) stops, inbound and outbound. In sum, there were 10 separate transit networks created to enable spatially and temporally disaggregate analysis in this study.

The spatial unit for measuring accessibility is the TAZ. There are four assumptions: (1) all the population living in the TAZ can access the transit stop that is closest to the centroid of that TAZ; (2) as the transit network is separated into inbound and outbound, population in origin TAZ can access one stop each in the inbound and outbound networks; (3) accessibility impacts are calculated for inbound and outbound networks separately, and the total accessibility impacts experienced by a TAZ during a specified time period is the sum of accessibility impacts via the inbound and outbound networks; (4) the wait time at the stops and the walk time to the stops are not considered in this study, since the analysis only considers travel between different TAZs (and not trips within a TAZ).

There are two types of accessibility measured in this study: work-related accessibility (based on employment) and non-work-related accessibility (based on POIs). For both measures, only TAZs with non-zero total population within a 0.25-mile radius buffer of a transit stop are considered as origin TAZs. Similarly, only TAZs with non-zero employment or non-zero count of POIs and fall within a 0.25-mile radius buffer of a transit stop are used for destinations. There are transit-accessible 710 TAZs (out of 1,173 total TAZs) in the Hampton Roads region based on this criterion. All trip origins and destination are selected from these TAZs. Based on the transit schedule and operating routes, the number of origin and destination vary for each time period, as shown in Table 2 in the Result section.

## 5.2 Accessibility Calculation

Gravity model with Gamma function was selected to calculate accessibility in this paper. For a detailed discussion of the gravity model and its associated parameters, refer to Section 4.1.1-4.1.2 of Part 1 of the study. For transit accessibility in this paper (Part 2 of the study), this accessibility calculation was applied to inbound and outbound trips separately for each TAZ, and the total accessibility of an individual TAZ will be the sum of them, as shown in equation (1).

$$A_{total} = A_{inbound} + A_{outbound} \tag{1}$$

Where:

    $A_{total}$:    the total accessibility of TAZ
    $A_{inbound}$:  the accessibility of TAZ in inbound transit network
    $A_{outbuond}$: the accessibility of TAZ in outbound transit network

The origin-destination (OD) matrix generation and calculations of change in accessibility remains the same as Section 4.1.3 in Part 1 of the study, as well as the social vulnerability index framework in Section 4.2 in Part 1 of the study, where SVI values (ranging from 0 to 1) closer to 1 indicate higher social vulnerability.

## 6. RESULTS AND DISCUSSION

Recurrent flooding-induced accessibility changes are assessed for transit-serving TAZs in the Hampton Roads region in this paper (Part 2). Figures 6(a) - 6(e) show the accessibility reduction ($\Delta A_i$) of each TAZ for work-related travel during each of the five time-of-day periods in August 2018. Similarly, Figures 7(a) - 7(e) show the $\Delta A_i$ for each TAZ for non-work-related travel as a result of recurrent flooding. TAZs with less than 1% reduction in accessibility are considered low impact (light green). TAZs with reduction in accessibility between 1 and 5%, inclusively, are considered medium impact (medium green). TAZs with reduction in accessibility above 5% are considered high impact (blue).

## 6.1 Transit Accessibility Impacts for the General Population

The geographical distinction between accessibility for work and non-work purposes is significant. During morning peak hour (6 a.m. – 9 a.m.). work accessibility is severely impacted in the City of Norfolk, Virginia Beach, and Chesapeake, as depicted by the blue TAZs in Figure 4(b). When it comes to non-work accessibility, the greatest impacts are concentrated at the southwest of Virginia Beach and southeast of Chesapeake as shown in Figure 5(b). In simpler terms, these findings indicate that during the morning peak hour, City of Norfolk experiences more significant impacts on work transit accessibility while Virginia Beach and Chesapeake faces greater impacts on both work and non-work transit accessibility.

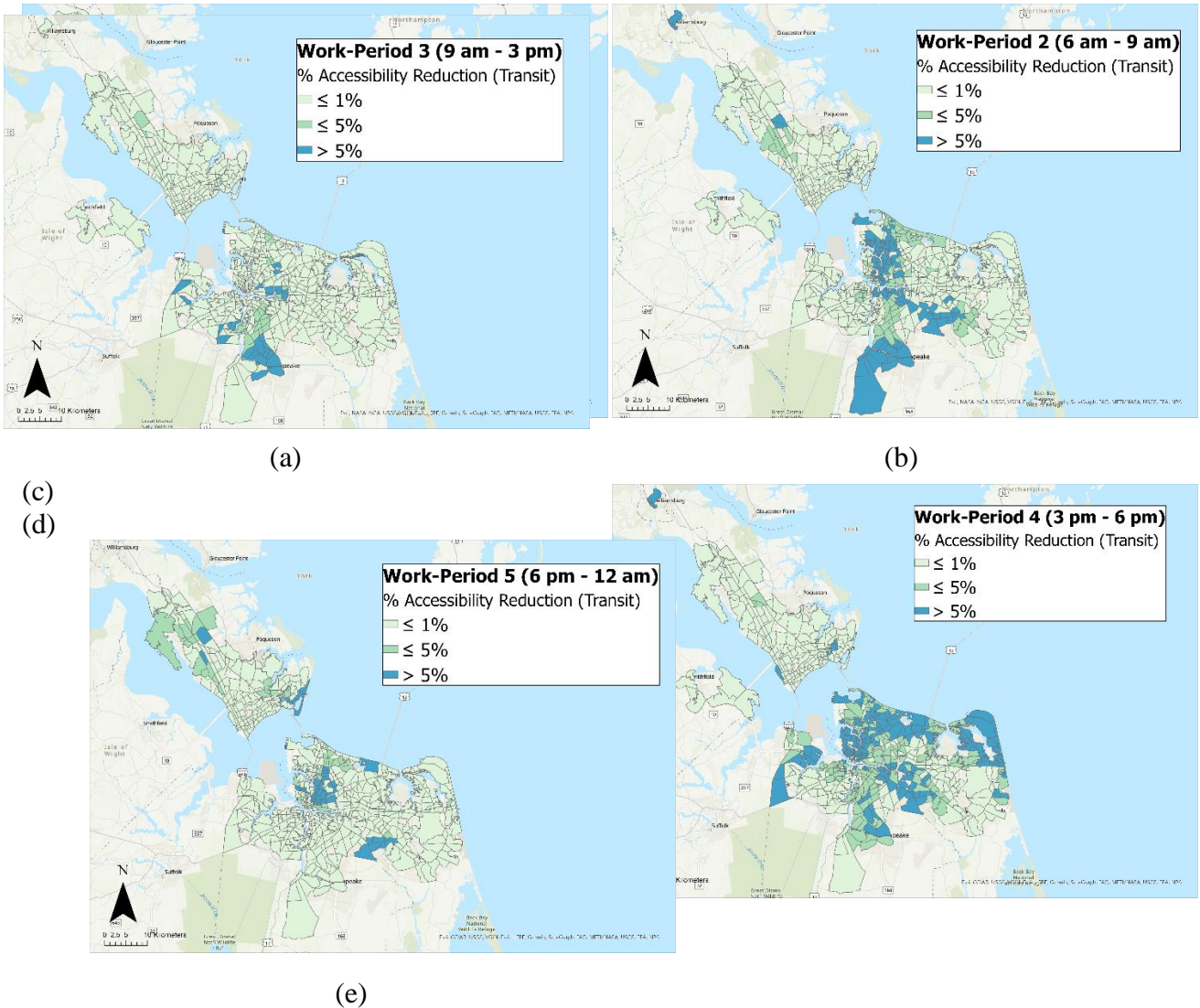

**Figure 4 Percent Change in Work Accessibility in (a) Period 1, (b) Period 2, (c) Period 3, (d) Period 4, and (e) Period 5**

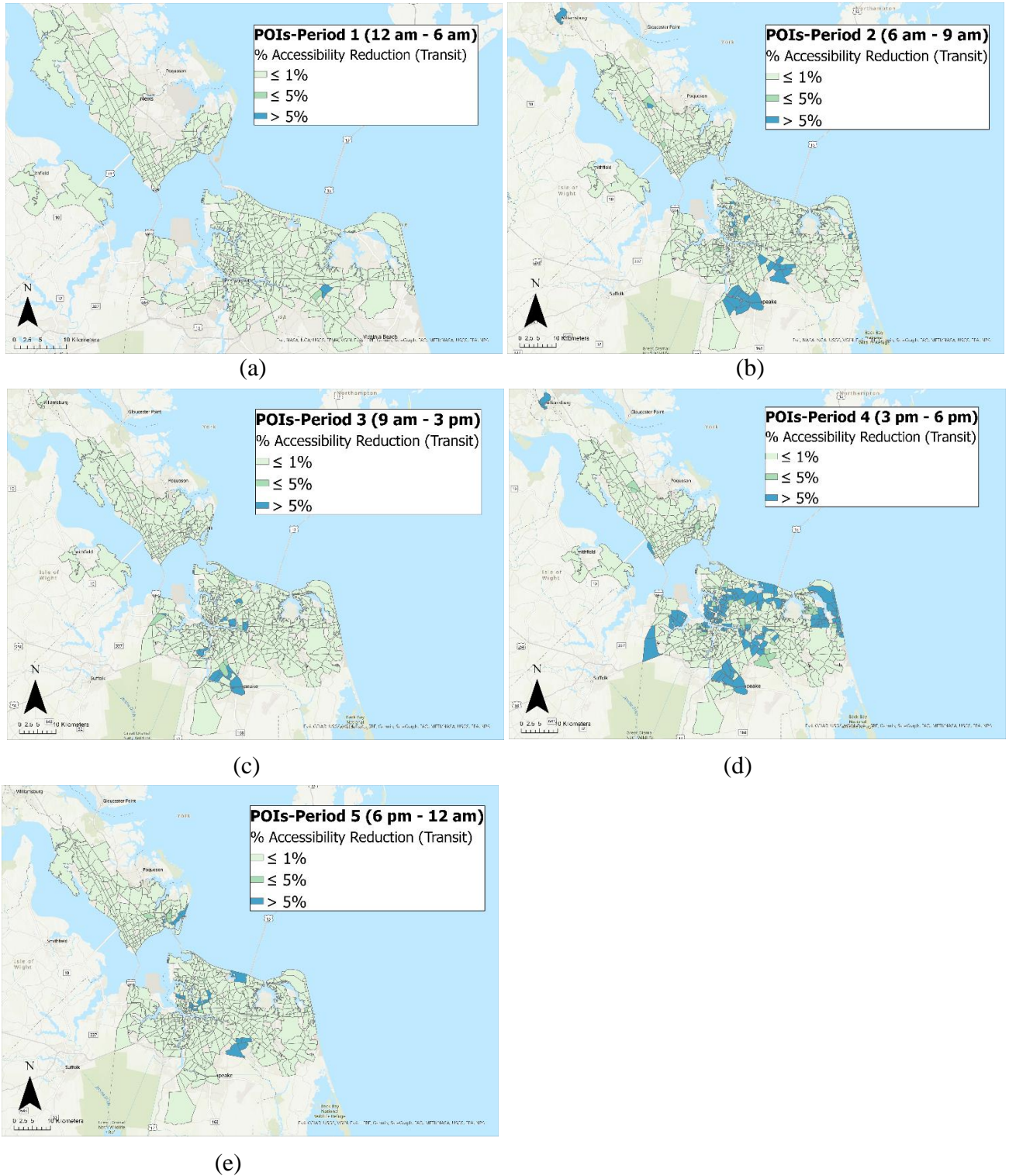

(a)

(b)

(c)

(d)

(e)

**Figure 5 Percent Change in POIs Accessibility in (a) Period 1, (b) Period 2, (c) Period 3, (d) Period 4, and (e) Period 5**

Table 2 summarizes the number of TAZs covered by transit service in each time period. Within the same time period, the number of TAZs covered for work and non-work trips are also different due to the different locations of the POIs. The distribution of TAZs in the three accessibility reduction categories over the five time-of-day periods for both work and non-work (point-of-interest) trips are also shown in Table 2.

**Table 2 Temporal Distribution of Accessibility Impacts**

| Destination Type | Time Period [a] | Number of TAZs served by transit service | % TAZs impacted | | | Maximum Accessibility Reduction in a single TAZ ($\Delta A\_i$) | Population weighted mean accessibility reduction across all TAZs |
|---|---|---|---|---|---|---|---|
| | | | Low Impact ($\leq$ 1%) | Medium Impact ($\leq$ 5%) | High Impact (>5%) | | |
| Work | 1 (midnight to 6 am) | 518 | 99.42% | 0.00% | 0.58% | 86.13% | 0.40% |
| | 2 (6 am to 9 am) | 663 | 67.12% | 13.88% | 19.00% | 88.16% | 6.42% |
| | 3 (9 am to 3 pm) | 663 | 91.55% | 3.93% | 4.52% | 31.73% | 0.82% |
| | 4 (3 pm to 6 pm) | 663 | 45.10% | 25.34% | 29.56% | 41.05% | 3.45% |
| | 5 (6 pm to midnight) | 650 | 83.85% | 8.00% | 8.15% | 25.33% | 1.08% |
| Non-Work | 1 (midnight to 6 am) | 511 | 99.61% | 0.00% | 0.39% | 84.82% | 0.37% |
| | 2 (6 am to 9 am) | 662 | 92.15% | 2.11% | 5.74% | 99.93% | 3.17% |
| | 3 (9 am to 3 pm) | 658 | 95.74% | 1.52% | 2.74% | 33.33% | 0.71% |
| | 4 (3 pm to 6 pm) | 660 | 73.33% | 5.61% | 21.06% | 40.76% | 2.56% |
| | 5 (6 pm to midnight) | 646 | 94.12% | 1.55% | 4.33% | 20.00% | 0.84% |

[a] Time periods include the beginning but not ending time point: a flood event that occurs at precisely 6:00:00 a.m. is attributed to period 2 rather than period 1.

For both work and non-work trips, period 2 (6 to 9 am) experiences the greatest transit accessibility reduction incurred by the recurrent flooding, with a maximum reduction of 88.2% and 99.9% in the most affected TAZs, respectively. Both these TAZs are located within the City of Norfolk, and it is important to recognize that select areas in the city lose transit accessibility almost completely under recurrent flooding. Meanwhile, the number of TAZs that experience more than 5% transit accessibility reduction (high impact) was the greatest for both work and non-work trips during Period 4 (3 to 6pm). As shown in Figures 6 (b) to (e) and Figures 7(b) to (e), although the opportunities and destinations are different for work and non-work tips, the highly impacted areas are relatively similar in spatial distribution in Periods 2 through 5, where most of them are within the City of Norfolk, Virginia Beach, and Chesapeake.

These results suggest that the accessibility impact of recurrent flooding varies by trip purpose and time of day. Of all transit-accessible TAZs, the population weighted accessibility reduction for work-related trips for highly impacted TAZ across all time periods was 36.7%, compared to 61.6% for non-work trips. Part of this disparity is due to the fact that the study focuses on fixed route transit service, and the location of work destinations are more evenly distributed across the transit-accessible TAZs, compared to non-work destinations which are exhibit more clustering. The spatial difference in work and non-work accessibility is also demonstrated by Figure 6, which show heat maps for how frequently some TAZs are highly impacted by recurrent flooding out of the 18 date-time period combinations.

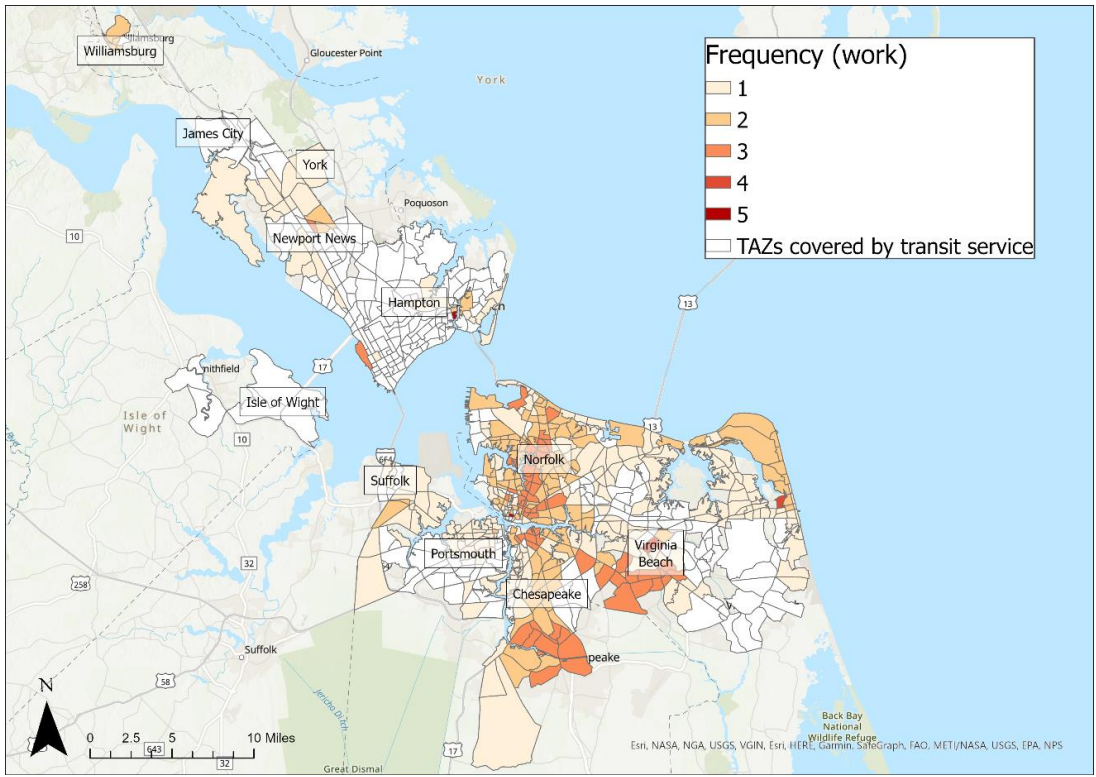

(a)

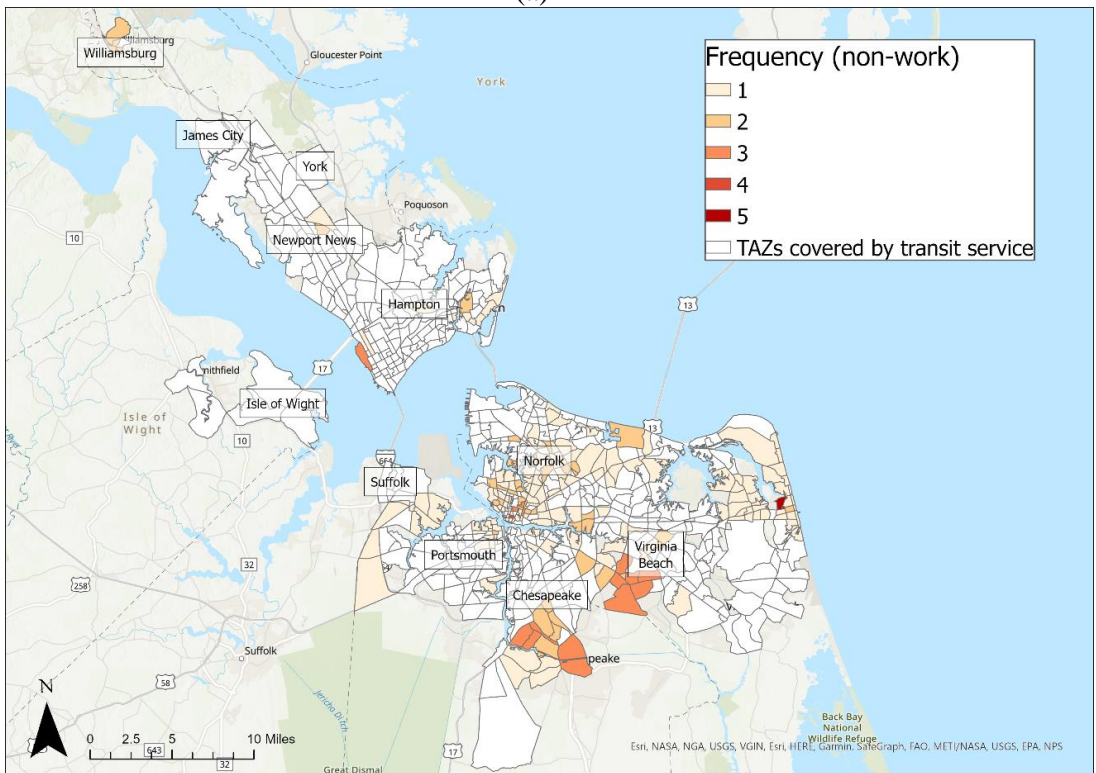

(b)

**Figure 6 Frequency of recurrent flooding impact of highly impacted TAZs for (a) work trips, (b) non-work trips.**

For work trips across the 8 days with flood reports in August 2018, 138 unique TAZs experienced significant flood-related accessibility reduction for more than one time period, whereas for non-work tips, 44 unique TAZs experienced more than 5% accessibility reduction for more than one period. As shown in Figure 8, the TAZs that experience the most frequent accessibility impact due to flood incidents are located within the cities of Norfolk and Hampton in close proximity to bodies of water, exacerbating roadway flooding with rainfall and high tide.

## 6.2 Transit Accessibility Impacts for the Vulnerable Population

Principal component analysis (PCA) is applied to the second order SVI for all the TAZs that experience high accessibility reduction under recurrent flooding for both trip purposes (refer to Part 1 section 5.2 for detailed discussion). The relationship between second order SVI variables (as shown in Figure 5) and transit accessibility reduction for each TAZs is tested. Results show that for both trip purposes, only the first two PCs have eigenvalue greater than 1 across 6 PCs, and the accumulated amount of explained variance is 66.4% for work trips and 68.1% for non-work trips, respectively. Therefore, for both trip purposes, only two PCs are retained, and the bi-plot is shown in Figure 7.

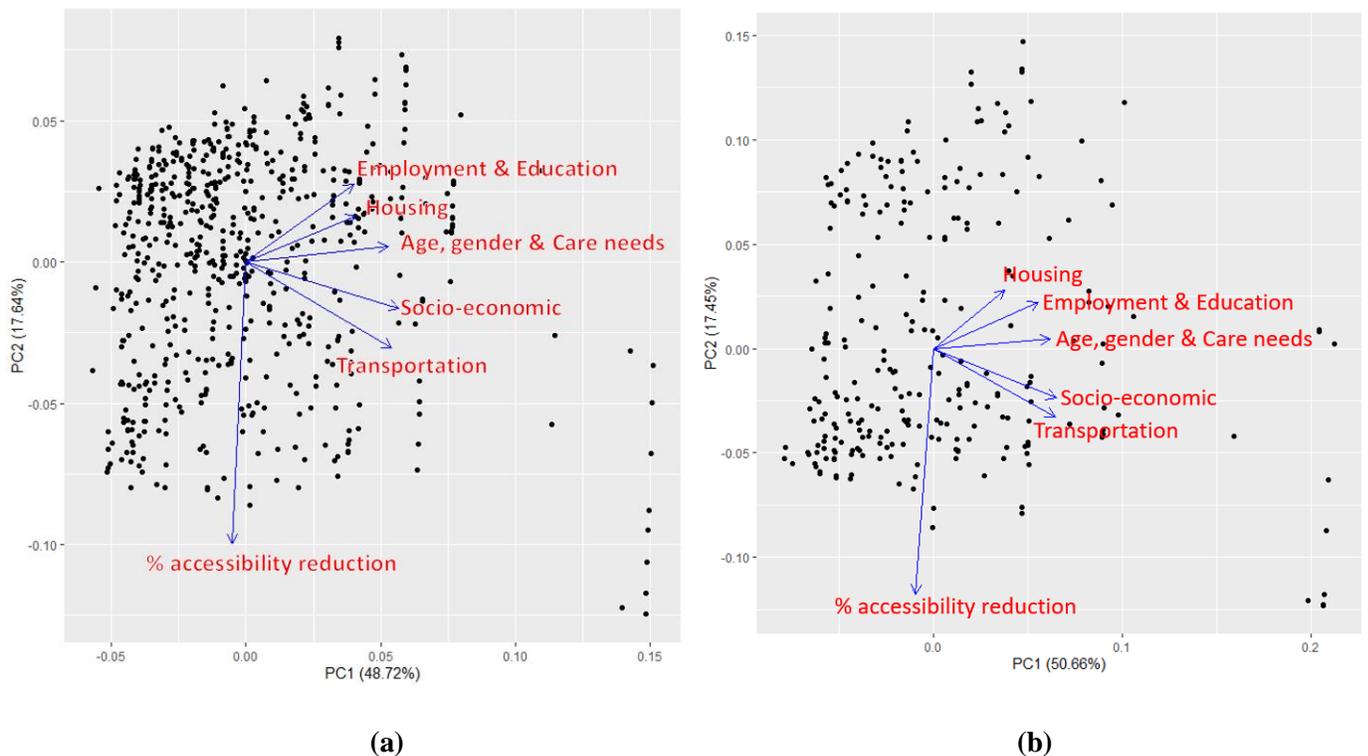

**(a)**                                                                          **(b)**

**Figure 7 PCA biplot for (a) work trips; (b) non-work trips**

For both work and non-work trips, as shown in Figure 9, populations experiencing higher vulnerability in transportation (i.e., those who commute to work via public transportation and are without a vehicle) and socio-economic status (e.g., lower income) are more likely to experience higher transit accessibility reduction caused by recurrent flooding. This finding again underscores the importance of recognizing the heterogeneous impacts of recurrent flooding on transportation. In this case, the population who relies most on transit and have lower income is also the group whose transit accessibility is affected the most by recurrent flooding. The relationship between third order SVI and transit accessibility reduction was examined with scatter plots. However, the correlation was weak, thus no conclusion can be drawn at this point.

## 7. CONCLUSIONS

The frequency of recurrent flooding events is increasing with accelerated sea level rise, and the impact of such events on transportation accessibility is not yet well understood. Using the Hampton Roads Regions as a case study, this paper (Part 2 of the study) investigates the spatial and temporal variation of transit accessibility reduction under recurrent flooding with the aid of GIS network analysis and crowdsourced flood incident and congestion reports. The study also demonstrates the usage of a modified CDC SVI framework to understand the socially vulnerable groups who are most susceptible to transit accessibility reduction under recurrent flooding.

Results suggest the greatest transit accessibility reduction occurs during the morning peak period (6 – 9 am) for both work and non-work trips, which is the same for auto accessibility as presented in Part 1 of the study. Though the average accessibility reduction across all TAZs is between 2.6% and 6.4% for the region, the impact is not uniformly distributed. Some TAZs practically lose all transit accessibility during recurrent flooding while other TAZs experience negligible impacts. Social vulnerability analysis results suggest that people with higher vulnerability in transportation (e.g., commute with public transit) and socio-economic status are more likely to experience greater transit accessibility reduction, for both work and non-work trips. These results highlight the disproportionate impacts of climate events borne by socially vulnerable groups (e.g., transit-reliant populations experiencing greater transit access reduction), and reaffirm the need to assess these impacts on a geographically disaggregate scale for policy-making.

This study demonstrated a framework for evaluating the impact of recurrent flooding on transit accessibility using commonly available data, and can be adapted to other locations. The spatial-temporal disaggregate data facilitate a focus on recurrent flooding's dynamic impacts on the transportation system, which is distinct from conventional static identification of areas with high flood risk. With fixed route transit, the quantified accessibility reduction reflects the impact variety in different zones across different service periods, which allows local governments to prioritize limited public resources on the basis of transit accessibility, and lead to improvement in public transportation service performance and advance social equity.

The paper has similar limitation as Part 1 of the study. Specifically, the study is limited by the temporal coverage of WAZE data, and did not weigh the attractiveness of destinations based on the size or importance of such destinations. Moreover, the transit network is hard coded as inbound and outbound, which means that trips requiring a transfer from an inbound route to an outbound route cannot be properly captured in the analysis. Although ArcGIS Pro provides a public transit evaluator for processing GTFS data, this tool is not sensitive to disruption and functions purely based on a pre-set schedule, and could not be used in this analysis. In future studies, with higher quality of network GIS data and automation of processing WAZE congestion data, the spatial and temporal coverage of the analysis framework could be expanded and applied to more regions that experience recurrent flooding. The identification of recurrent flooding hot spots and potential social vulnerable populations in those zones could inform prioritization of mitigation measures and relevant policies.

**ACKNOWLEDGMENTS**
The authors thank the Hampton Roads Transportation Planning Organization, transitland, and WAZE for facilitation of data acquisition. The authors would also like to thank Afrida Raida, Linda Lim, and Andrea Saglio for giving valuable input in reviewing the manuscript. This work is supported by the National Science Foundation's Critical Resilient Interdependent Infrastructure Systems and Processes program (Award 1735587).

**AUTHOR CONTRIBUTIONS**
The authors confirm contribution to the paper as flows: study conception and design: L. Zeng, T.D. Chen, J. Miller, F.T. Zahura, J.L. Goodall; data collection and processing: L. Zeng, F.T. Zahura; analysis and interpretation of result: L. Zeng, T.D. Chen, J. Miller; draft manuscript preparation: L. Zeng, T.D. Chen. All authors reviewed and approved the final version of the manuscript.